\documentclass[twocolumn,aps,prl]{revtex4-1}
\usepackage{graphicx}
\usepackage{amsmath}
\usepackage{natbib}

\begin{document}

\title{Breaking of valley degeneracy by magnetic field in monolayer MoSe$_2$}
\author{David MacNeill}
\affiliation{Department of Physics, Cornell University, Ithaca, NY 14853, USA}
\author{Colin Heikes}
\affiliation{Department of Physics, Cornell University, Ithaca, NY 14853, USA}
\author{Kin Fai Mak}
\affiliation{Department of Physics, Cornell University, Ithaca, NY 14853, USA}
\affiliation{Kavli Institute at Cornell, Cornell University, Ithaca, NY 14853, USA}
\author{Zachary Anderson}
\affiliation{Department of Physics, Cornell University, Ithaca, NY 14853, USA}
\author{Andor Korm\'anyos}
\affiliation{Department of Physics, University of Konstanz, D-78464 Konstanz, Germany}
\author{Viktor Z\'olyomi}
\affiliation{Department of Physics, Lancaster University, Lancaster LA1 4YB, United Kingdom}
\author{Jiwoong Park}
\affiliation{Department of Chemistry and Chemical Biology, Cornell University, Ithaca, NY 14853, USA}
\affiliation{Kavli Institute at Cornell, Cornell University, Ithaca, NY 14853, USA}
\author{Daniel C. Ralph}
\affiliation{Department of Physics, Cornell University, Ithaca, NY 14853, USA}
\affiliation{Kavli Institute at Cornell, Cornell University, Ithaca, NY 14853, USA}
\date{{\small \today}}

\begin{abstract}
Using polarization-resolved photoluminescence spectroscopy, we investigate breaking of valley degeneracy by out-of-plane magnetic field in back-gated monolayer MoSe$_2$ devices.  We observe a linear splitting of $-0.22 \frac{\text{meV}}{\text{T}}$ between luminescence peak energies in $\sigma_{+}$ and $\sigma_{-}$ emission for both neutral and charged excitons. The optical selection rules of monolayer MoSe$_2$ couple photon handedness to the exciton valley degree of freedom, so this splitting demonstrates valley degeneracy breaking.  In addition, we find that the luminescence handedness can be controlled with magnetic field, to a degree that depends on the back-gate voltage. An applied magnetic field therefore provides effective strategies for control over the valley degree of freedom.
\end{abstract}
\maketitle

Monolayer MoSe$_2$ and other monolayer transition metal dichalcogenides (TMDs) are a materials system with unique potential for controlling their valley degree of freedom \cite{PhysRevLett.108.196802,mak2012control, jones2013optical, kioseoglou2012valley, zeng2012valley, cao2012valley, PhysRevB.86.081301, mak2014valleyhall}. Similar to graphene, the conduction and valence band show extrema (valleys) at the vertices of a hexagonal Brillouin zone; unlike graphene, MoSe$_2$ exhibits a nonzero optical gap of 1.66 eV  \cite{ross2013electrical, zhang2013direct}. This has allowed exploration of optoelectronic properties arising from the valley-dependent chirality of massive Dirac fermions, predicted in the context of inversion symmetry broken graphene \cite{PhysRevLett.99.236809, PhysRevB.77.235406}. This chirality leads to optical selection rules coupling the exciton valley degree of freedom to photon handedness  \cite{mak2012control, jones2013optical, kioseoglou2012valley, zeng2012valley, cao2012valley, PhysRevB.86.081301}. Using polarization-resolved spectroscopy researchers have demonstrated valley-selective luminescence with near 100$\%$ fidelity \cite{mak2012control, PhysRevB.86.081301}. Furthermore, the ability to pump valley-polarized carriers with circularly-polarized light has been demonstrated through the valley Hall effect \cite{mak2014valleyhall}. The chiral electronic states are also predicted to posses valley-contrasting orbital magnetic moments coupling valley pseudospin to magnetic field \cite{PhysRevLett.99.236809, PhysRevB.77.235406,PhysRevX.4.011034,PhysRevLett.110.066803, PhysRevB.90.045427, PhysRevB.88.085440,PhysRevB.89.155316}, which opens up the possibility for magnetic control over the valley degree of freedom  \cite{PhysRevB.88.115140,PhysRevX.4.011034}.  

Here, we demonstrate the use of magnetic fields to break valley degeneracy in a monolayer TMD. Specifically, we report polarization-resolved luminescence spectra for back-gated MoSe$_2$ devices at 4.2 K and in magnetic fields up to 6.7 T. We study the luminescence peak energies as a function of magnetic field, finding a linear splitting of $-0.22 \frac{\text{meV}}{\text{T}}$ between peaks corresponding to light emission with different senses of circular polarization, $\sigma_{+}$ and $\sigma_{-}$. We interpret this as a Zeeman splitting due to valley-dependent magnetic moments. We also investigate the magnetic field dependence of luminescence handedness, finding that the emission becomes circularly-polarized in magnetic field even with unpolarized excitation, and that the degree of this polarization can be increased to about $50 \%$ by gating the sample. This suggests that electric fields can facilitate the generation of valley-population imbalance in samples where valley degeneracy has been broken by magnetic field. Our results demonstrate a recently-proposed \cite{PhysRevB.88.115140} strategy for generating valley populations, and could lead to new approaches for controlling the valley degree of freedom in monolayer TMDs. 

\begin{figure}
\begin{center}
\includegraphics[width=\columnwidth]{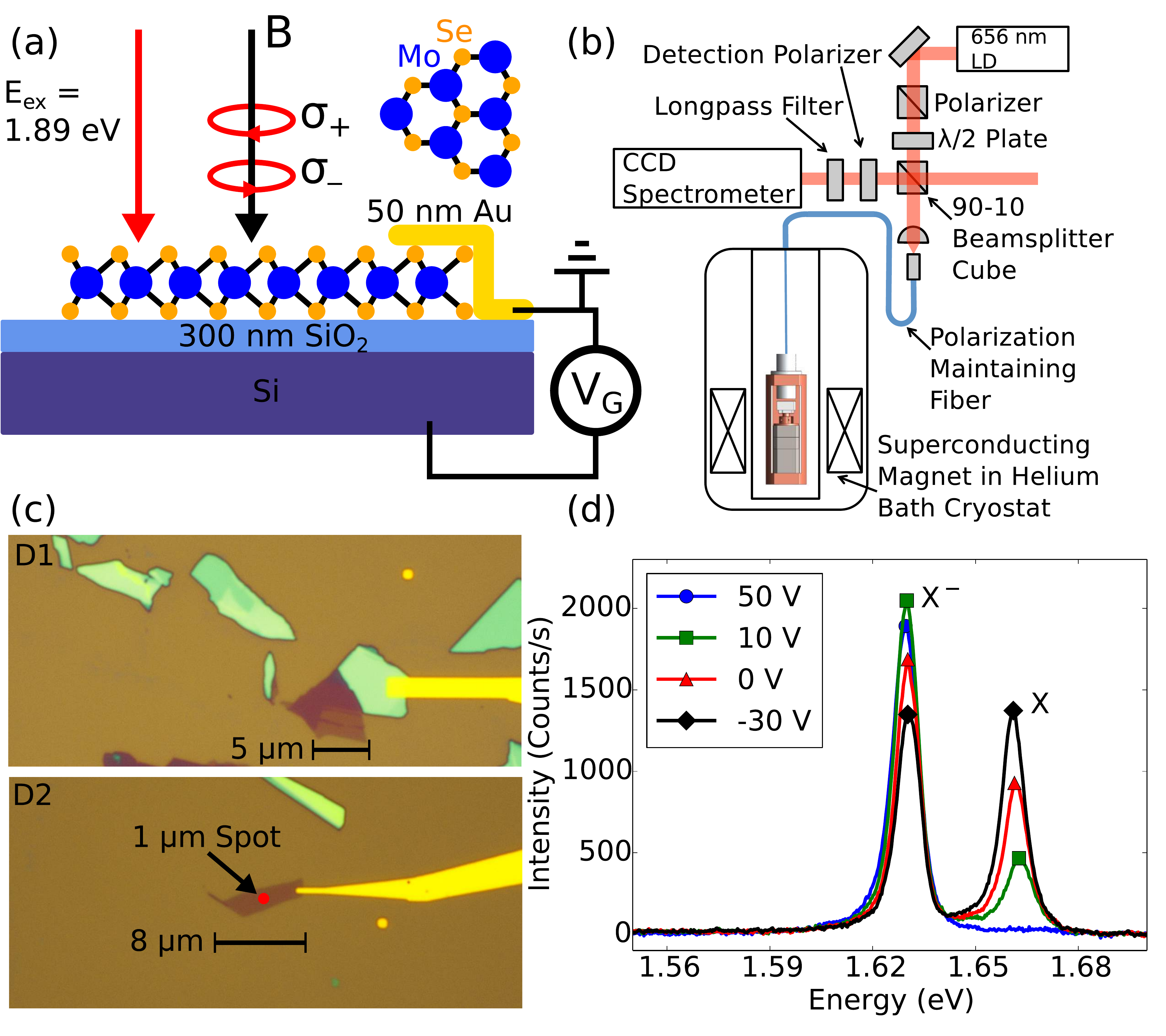}
    \caption{ (color online). (a) Experimental geometry showing back-gated monolayer MoSe$_2$ devices in out-of-plane magnetic fields. Luminescence is excited with light from a 1.89 eV laser diode and collected separately for $\sigma_+$ and $\sigma_-$ polarization in the Faraday geometry. (b) Schematic of the fiber-coupled optical cryostat used in the experiment. (c) Optical micrographs of devices D1 and D2. (d) Luminescence spectra of D2 taken at 0 T and 4.2 K with -30 V, 0 V, 10 V, and 50 V back-gate voltage.} \label{fig1}
\end{center}
\end{figure}

Our device geometry and measurement apparatus are shown in Fig.\ \ref{fig1}a and \ref{fig1}b. All measurements were taken using a scanning confocal microscope integrated with a 7 T superconducting magnet dewar, with light coupled in and out of the system via a polarization-maintaining optical fiber (similar designs were reported in Refs.\ \cite{hogele2008fiber, sladkov2011polarization}). The light is focused into a roughly 1 $\mu$m diameter spot using a pair of aspheric lenses, and the sample is scanned using piezo-driven nanopositioners (from attocube). The sample, positioners, and optical components are placed in a vacuum cryostat which is then evacuated and lowered into a helium bath containing a superconducting magnet; helium exchange gas is added to ensure thermalization of the sample at 4.2 K. For the data in the main text, the excitation power was between 10-60 $\mu$W. 

To enable polarization-resolved spectroscopy, a zero-order quartz $\lambda/4$ plate is placed between the aspheric lenses, oriented at 45$^{\circ}$ to the fiber axes; this couples $\sigma_{+}$ and $\sigma_{-}$ emission into orthogonal polarization modes of the fiber. The light exiting the fiber is directed though a rotatable polarizer, which selects one fiber mode for spectral analysis by a thermoelectrically cooled CCD spectrometer. We can also create circularly-polarized excitation by coupling linearly-polarized light into one of the two fiber polarization modes, or create equal intensity excitation in $\sigma_+$ and $\sigma_-$ polarization by coupling in light polarized at 45$^\circ$ to the fiber axes. We excite photoluminescence with light from a 1.89 eV laser diode, which is 230 meV blueshifted from the A exciton transition, and as a result we see little dependence of the emission polarization on excitation polarization (see supplement section 1). The conclusions discussed below are independent of excitation polarization. 

To fabricate our samples, we exfoliate bulk MoSe$_2$ crystals (grown by direct vapor transport) onto 300 nm silicon oxide on silicon, then use electron-beam lithography to define a single 0.5 nm Ti/75 nm Au contact, allowing use of the silicon substrate as a back gate. All data shown in the main text were taken from devices D1 and D2 pictured in Fig.\ \ref{fig1}c. Figure \ref{fig1}d shows the $B=0$ luminescence spectra of D2 at -30 V, 0 V, 10 V, and 50 V. The peaks at 1.66 eV and 1.63 eV correspond to the neutral and charged A exciton respectively, with a charged exciton (trion) binding energy of 30 meV \cite{ross2013electrical}. As the back-gate voltage is increased the exciton luminescence decreases and the trion luminescence increases, showing that our samples are intrinsically $n$-type and that the 1.63 eV peak corresponds to negatively charged trion luminescence. 
\begin{figure}
\begin{center}
\includegraphics[height=0.5\textheight]{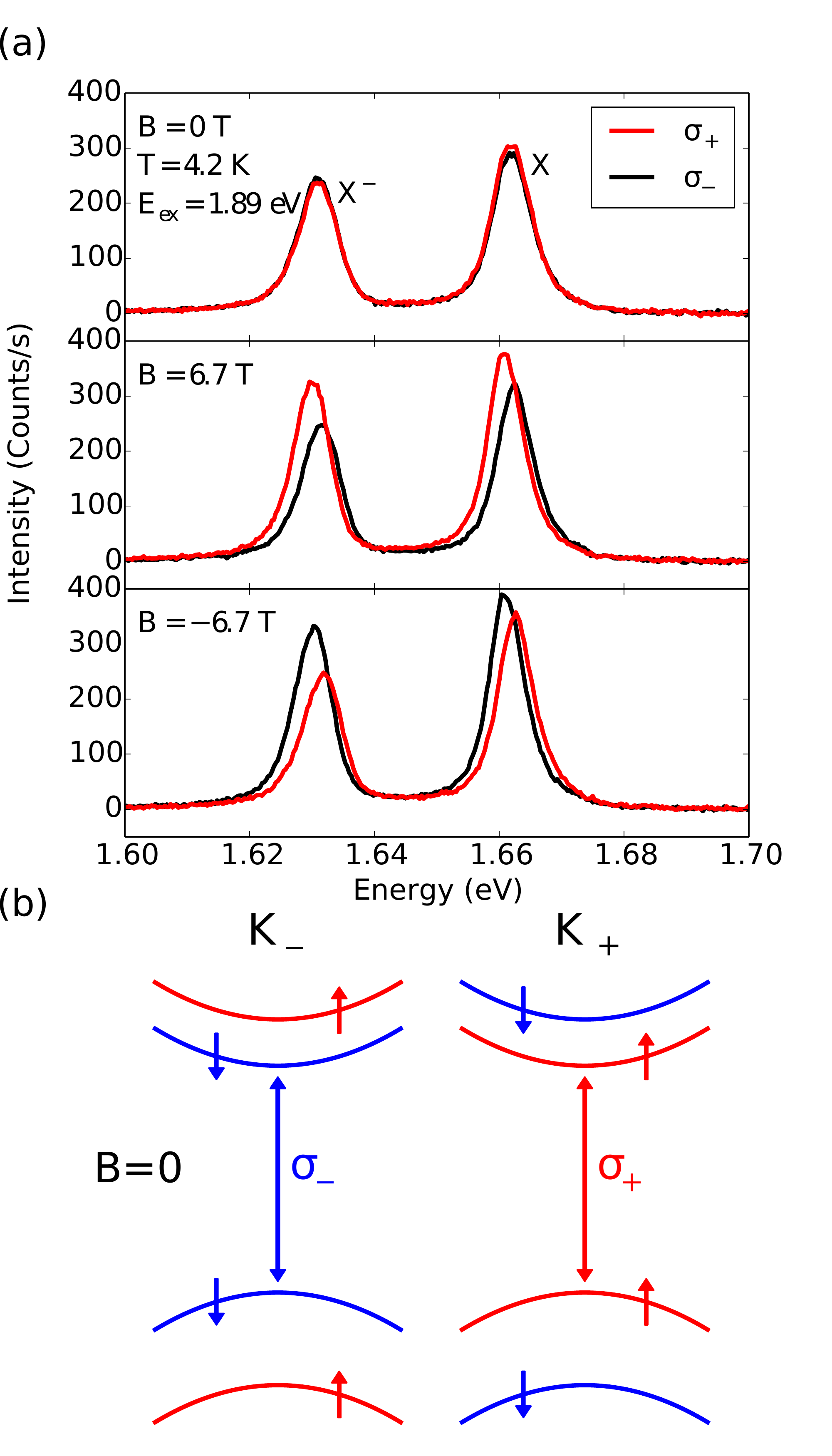}
    \caption{(color online). (a) Polarization-resolved luminescence spectra from monolayer MoSe$_2$ (D1) at 4.2 K for $\sigma_{+}$ and $\sigma_{-}$ detection, as excited using unpolarized light at 1.89 eV. From top to bottom the panels show spectra taken with 0 T, 6.7 T and -6.7 T out-of-plane magnetic field. Both the polarization and splitting change sign upon reversing the field as shown in the lower panel. (b) Schematic bandstructure of MoSe$_2$ near the $K_+$ and $K_-$ points in zero magnetic field, showing the optical selection rules for the A exciton transition studied in this experiment. Within each valley, spin degeneracy is broken at $B=0$ due to spin-orbit coupling, \cite{zhang2013direct, ross2013electrical, PhysRevX.4.011034, PhysRevB.88.085433,PhysRevB.88.245436}. The arrows denote spin angular momentum up and down for the occupied states. } \label{fig2}
\end{center}
\end{figure}

Figure \ref{fig2}a compares polarization-resolved spectra taken for D1 in out-of-plane magnetic fields of 0 T, 6.7 T and -6.7 T and with the back gate grounded. For these data, we excite photoluminescence using equal intensity excitation in $\sigma_+$ and $\sigma_-$ polarization. At zero field, we find no significant dependence of the peak energies or intensities on emission handedness. In comparison, the spectra taken at 6.7 T show splitting between the $\sigma_{+}$ and $\sigma_{-}$ emission peaks of about -1.5 meV for both the exciton and trion. The luminescence is also $\sigma_{+}$ polarized: the trion peak has $P_{\text{trion}}=\frac{I_{+}-I_{-}}{I_{+}+I_{-}} =14\% $, where $I_{\pm}$ is the peak intensity of the trion in $\sigma_{\pm}$ detection. For the exciton we measure $P_{\text{exciton}}=9\%$. The luminescence polarization changes sign with reversal of the magnetic field but not with excitation polarization, showing that it arises from magnetically induced changes in the exciton and trion populations. Figure \ref{fig2}b depicts the schematic bandstructure of a MoSe$_2$ monolayer, illustrating the direct band gaps at the $K_+$ and $K_-$ points, with arrows indicating the allowed A exciton transitions for $\sigma_{\pm}$ light. Since the emission handedness is coupled to the exciton valley degree of freedom, the peak splitting and polarization we observe indicate valley degeneracy breaking. 

\begin{figure}
\begin{center}
\includegraphics[height=0.5\textheight]{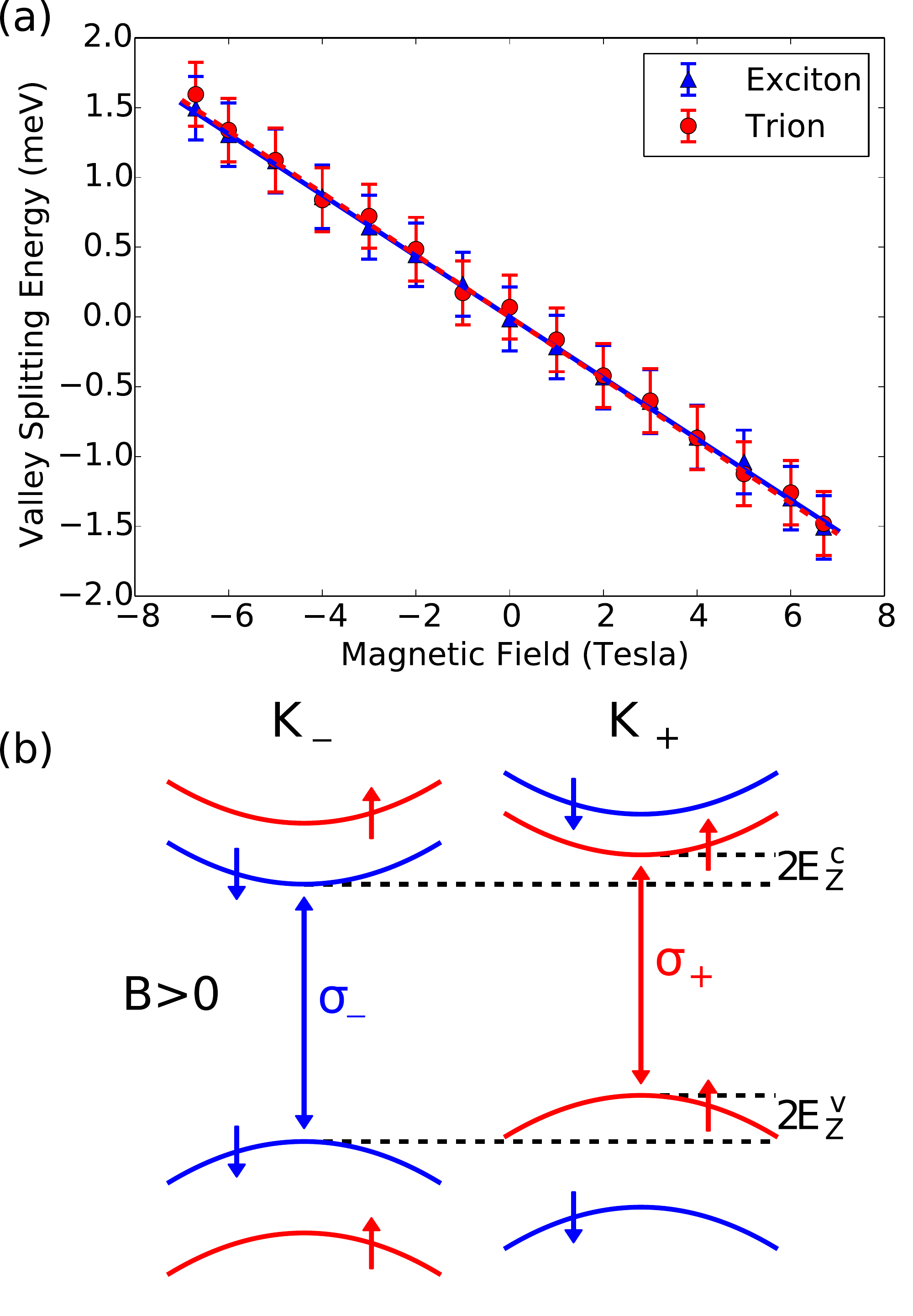}
\end{center}
\caption{(color online) (a) Difference of peak energies found for $\sigma_{+}$ and $\sigma_{-}$ detection plotted versus magnetic field for D1.  Both the exciton (blue triangles) and trion (red circles) show splitting of $-0.22 \pm 0.01 \frac{\text{meV}}{\text{T}}$ found via a linear fit. The fits are plotted as blue solid and red dashed lines for the exciton and trion respectively. (b) The schematic bandstructure of MoSe$_2$ in magnetic field showing the Zeeman energy $E_Z^{c(v)}$ for the conduction (valence) band. The exciton Zeeman splitting is $2\left(E_Z^{c}-E_Z^v\right)$. } \label{fig3}
\end{figure}

Figure \ref{fig3}a shows the valley splitting of the exciton and trion peaks, defined as the difference between peak luminescence energy in $\sigma_{+}$ and $\sigma_{-}$ detection, versus magnetic field. For each data point the peak positions were extracted via fits to a phenomenological asymmetric Voigt line shape (see supplement section 2). The errorbars come primarily from the CCD pixel size (about 0.15 nm per pixel). For both the exciton and trion peaks the valley splitting shows a linear magnetic-field dependence with a slope of $-0.22 \pm 0.01 \, \frac{\text{meV}}{\text{T}}$, and consistent results were found on three separate devices; data from other devices are given in supplement section 3. 

Valley splitting in magnetic field arises from the intrinsic chirality of Bloch electrons at the $K_+$ and $K_-$ points. States at the two valley edges are Kramer's doublets related by time-reversal symmetry, so that their degeneracy can be broken by breaking time-reversal symmetry. Bloch electrons in a given band carry spin and orbital magnetic moments which change sign between valleys \cite{PhysRevB.77.235406,PhysRevLett.99.236809,0953-8984-20-19-193202, PhysRevB.88.045416}. Figure \ref{fig3}b schematically shows the energy shifts arising from Zeeman coupling between these moments and the magnetic field; there we define $2E_Z^{c(v)}$ as the magnetic-field-induced energy difference between the $K_{+}$ and $K_{-}$ valley at the conduction (valence) band edge.  Magnetoluminescence spectroscopy probes only the exciton Zeeman energy, which is the difference between conduction and valence band Zeeman energies. In this difference, the contributions from spin magnetic moments are expected to cancel, leaving only the contributions from orbital magnetic moments. The measured sign and magnitude of the valley splitting can be understood within a tight-binding picture \cite{srivastava2014valley, aivazian2014magnetic}. In the $K_{\tau}$ valley (letting $\tau=\pm1$ be the valley quantum number), the valence band arises from hybridization of $d_{x^2-y^2}+\tau id_{xy}$ orbitals with angular momentum $l_z=2 \tau \hbar$ while the conduction band arises from hybridization of $d_{z^2}$ orbitals with $l_z=0$ \cite{PhysRevLett.108.196802, cao2012valley, zhu2011giant, PhysRevB.88.245436}. In the tight-binding limit, we therefore expect a contribution to the exciton Zeeman energy of $2\left(E_{Z,\text{a}}^{c}-E_{Z,\text{a}}^v\right)=-4\mu_{\text{B}}B$ from atomic-scale magnetic moments. The phase winding of Bloch states on the intercellular scale can also add to the orbital magnetic moment \cite{srivastava2014valley, aivazian2014magnetic, PhysRevLett.99.236809, 0953-8984-20-19-193202, yafet1963solid}. For example, in the two-band tight-binding model (the massive Dirac fermion model) the intercellular magnetic moment is equal for the conduction and valence bands with value $-\tau\mu_{\text{B}}\frac{m_e}{m_{\text{eff}}}$, where $m_e$ is the free-electron mass, and $m_{\text{eff}}$ is the electron-hole symmetric carrier effective mass \cite{PhysRevB.77.235406, PhysRevLett.99.236809}. Including the spin magnetic moments this gives a total Zeeman splitting of $2E_Z^{c}=2\mu_{\text{B}}+2\mu_{\text{B}}\frac{m_e}{m_{\text{eff}}}$ for the conduction band and $2E_Z^{v}=2\mu_{\text{B}}B+4\mu_{\text{B}}B+2\mu_{\text{B}}B\frac{m_e}{m_{\text{eff}}}$ for the valence band, and as a result $2\left(E_Z^{c}-E_Z^v\right)=-4\mu_{\text{B}}B$ (i.e. there is no net intercellular contribution). In more general hopping models, the conduction and valence bands can have different intercellular moments giving a net contribution to the exciton magnetic moment \cite{srivastava2014valley, aivazian2014magnetic, PhysRevB.88.085433, PhysRevB.88.085440}. 

To compare our measurements with theory, we define the exciton valley g-factor $g^{\text{vl}}_{\text{ex}}$ as:
\begin{equation}\label{eqn1}
g^{\text{vl}}_{\text{ex}}=\frac{E_+-E_-}{\mu_{\text{B}}B}=\frac{2(E_Z^{c}-E_Z^v)}{\mu_{\text{B}}B}
\end{equation} 
where $E_{\pm}$ is the measured exciton peak energy in $\sigma_{\pm}$ detection. Our exciton valley splitting measurements correspond to $g^{\text{vl}}_{\text{ex}}=-3.8\pm0.2$, consistent with the value of $g^{\text{vl}}_{\text{ex}}=-4$ expected from the $d$-orbital contribution to the exciton magnetic moment. Any deviation of $g^{\text{vl}}_{\text{ex}}$ from $-4$ theoretically corresponds to the intercellular contribution to the g-factor. Our results therefore suggest  that the  intercellular contribution to $g^{\text{vl}}_{\text{ex}}$ is small in the case of MoSe$_2$. We also expect the trion to have approximately the same splitting as the exciton, evinced by considering the trion as an exciton bound to an additional electron. While the additional electron contributes to the trion magnetic moment, it contributes equally to the final state moment after recombination leaving the transition energy unaffected (see supplement section 4). This is consistent with the experimental results of Fig.\ \ref{fig3}a for zero applied gate voltage. 

We also attempted to calculate the valley g-factor using the multiband $\mathbf{k\cdot p}$ theory of Refs.\ \cite{PhysRevB.88.045416, PhysRevX.4.011034}, since their theory should include the intercellular and atomic contributions in a unified way \cite{yafet1963solid}. The need to discuss these terms separately is an artifact of the lattice models discussed above. The calculation is detailed in section 5 of the supplement and gives a value for $g^{\text{vl}}_{\text{ex}}$ similar in magnitude to our experimental results, but with the opposite sign (see supplement section 6 for our experimental determination of the sign). Therefore further theoretical work is required to understand the exciton valley splitting within the context of $\mathbf{k\cdot p}$ theory calculations.

\begin{figure}
\begin{center}
\includegraphics[width=\columnwidth]{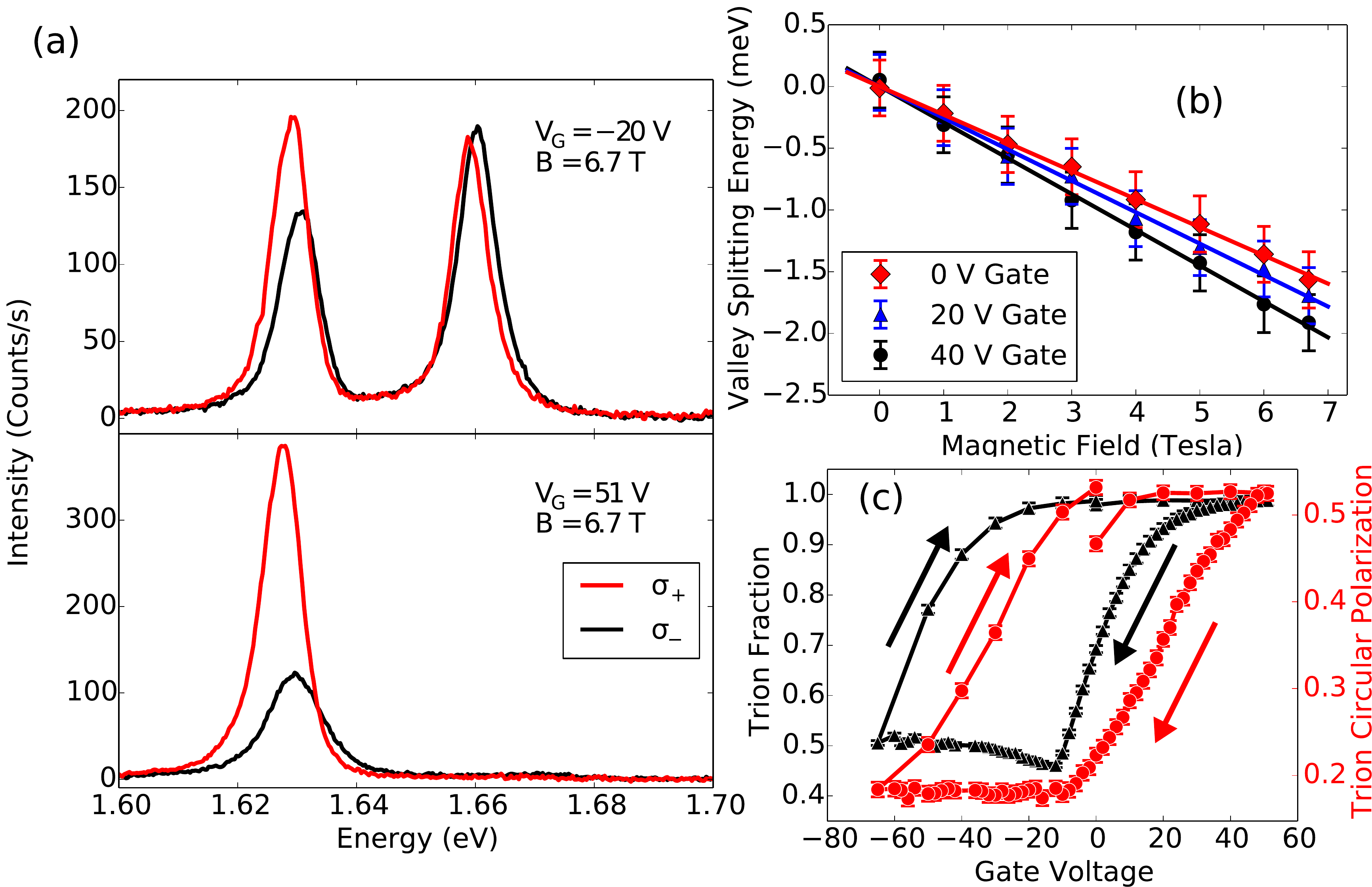}
\end{center}
\caption{(color online). (a) Polarization-resolved luminescence spectra from D2 at 4.2K and 6.7 T for $\sigma_{+}$  and $\sigma_{-}$ detection, excited with $\sigma_{-}$ light at 1.89 eV. From top to bottom the panels show spectra taken with -20 V and 51 V gate voltage applied to the substrate. (b) Valley splitting versus magnetic field for selected gate voltages, showing an decrease in slope with gate voltage. (c) Circular polarization of the trion peak $\frac{I_{+}-I_{-}}{I_{+}+I_{-}}$ versus gate voltage at 6.7 T (red circles), showing an increase to over $50\%$ as gate voltage is increased. For comparison, we also plot the trion fraction $\frac{I_{\text{trion}}}{I_{\text{trion}}+I_{\text{exciton}}}$  (black triangles). }\label{fig4}
\end{figure}

We find that the valley splitting and the resulting luminescence polarization both show a surprising dependence on an applied back-gate voltage. Polarization-resolved spectra taken with -20 V and 51 V applied to the substrate are shown in Fig.\ 4a for device D2. Our samples show significant hysteresis assumed to arise from photoionization of trap states \cite{PhysRevB.88.245403}, and the data in this panel are taken from a downward sweep. Figure 4b shows the trion splitting versus magnetic field for two different gate voltages on a downward sweep, finding $-0.29\pm0.02 \, \frac{ \text{meV}}{\text{T}}$ at 40 V and $-0.23\pm0.02 \, \frac{ \text{meV}}{\text{T}}$ at 0 V. This gate-voltage dependence of the valley splitting could arise from carrier-density dependence of the band Zeeman energies \cite{PhysRevB.88.085440, PhysRevLett.99.236809}, a hot luminescence effect as discussed in section 4 of the supplement, or other effects resulting from changes in the trion or final state wavefunctions upon increasing the Fermi level \cite{PhysRevB.89.205436}. The gate dependence of trion valley splitting has implications for future magneto-optical studies of TMDs, as the intrinsic doping level may vary between samples causing a dispersion of measurement results.

The degree of trion polarization as a function of gate voltage is shown in Fig.\ 4c. In this dataset, we find a trion polarization that increases from $18\%$ near zero back-gate voltage to over $50\%$ near 40 V.  The luminescence polarization is related to the valley population via $P_{\text{trion}}=\frac{n_{+}-n_{-}}{n_{+}+n_{-}}$, where $n_{\pm}$ is the trion population in valley $K_{\pm}$. From this we infer that we are observing the generation of valley-polarized trion populations through applied magnetic field and gate voltage. The sign of $P_{\text{trion}}$ in the $n$-type regime is independent of the excitation polarization but instead follows the sign of the magnetic field. We interpret the magnetic field dependence of the trion polarization as arising from partial relaxation of trions into their lowest energy spin-valley configuration (qualitatively consistent with the dependence of trion polarization on excitation power, see supplement section 7). This relaxation is expected to be incomplete as the intervalley scattering time is longer than the recombination time \cite{PhysRevB.89.205436, mak2012control}. In section 4 of the supplement, we calculate the trion polarization within a simple rate-equation model and show that the observed $P_{\text{trion}}$ implies a ratio of the recombination time to the intervalley scattering time of $\approx0.2$ at low carrier density. This is about an order of magnitude larger than the value found in time-resolved measurements for WSe$_2$ at zero magnetic field \cite{PhysRevB.89.205436}; however, the time-resolved measurements used resonant excitation which is expected to lead to reduced intervalley scattering compared to the off-resonant excitation we use. Trions can scatter between valleys via spin-flip intervalley scattering of their hole, and if this is the dominant scattering mechanism our results imply that the hole intervalley scattering rate increases monotonically with carrier density. This is consistent with the Bir-Aronov-Pikus mechanism for intervalley scattering of holes via their exchange interaction with the conduction electrons \cite{mak2012control, bir1975spin}. The data in Fig. \ref{fig4}c was taken with $\sigma_-$ excitation, but similar results were found using unpolarized excitation (see section 3 of the supplement).

In summary, we have presented measurements of polarization-resolved luminescence spectra for MoSe$_2$ at 4.2 K in magnetic fields up to 6.7 T, demonstrating valley degeneracy breaking. We measure a splitting of $-0.22\pm0.01\, \frac{ \text{meV}}{\text{T}}$ between exciton peaks in $\sigma_{+}$ and $\sigma_{-}$ polarized emission spectra.     
This value is consistent with a simple tight-binding picture of the MoSe$_2$ bandstructure. We also observe gate dependence of the valley splitting and polarization. Even with off-resonant, unpolarized excitation we were able to achieve a trion circular polarization of about 50$\%$ by gating the sample in 6.7 T magnetic field.  Application of magnetic and electric fields can therefore provide an effective strategy for manipulating the valley degree of freedom in monolayer TMDs. 

Similar work on WSe$_2$ has recently been posted by the Washington group \cite{aivazian2014magnetic} and the ETH Zurich group \cite{srivastava2014valley}.

We thank Kathryn McGill and Joshua Kevek for growth of the bulk MoSe$_2$ crystal used for this work. We also thank Guido Burkard, P\'eter Rakyta, Alexander H{\"o}gele and Ermin Malic for helpful discussions. This research was supported in part by the NSF (DMR- 1010768) and the Kavli Institute at Cornell for Nanoscale Science. We also made use of the Cornell Center for Materials Research Shared Facilities which are supported through the NSF MRSEC program (DMR-1120296). Device fabrication was performed at the Cornell NanoScale Facility, a member of the National Nanotechnology Infrastructure Network, which is supported by the National Science Foundation (Grant ECCS-0335765). D.\ M.\ acknowledges support from a NSERC PGS-D scholarship.  

\bibliography{ValleydegeneracybreakingbymagneticfieldinmonolayerMoSe2}

\onecolumngrid
\appendix

\section{{\large Supplement to ``Valley degeneracy breaking by magnetic field in monolayer MoSe$_2$"}}

\section{1. Dependence of Luminescence Handedness on Excitation Handedness}

\begin{figure*}[h]
\begin{center}
\includegraphics[width=\textwidth]{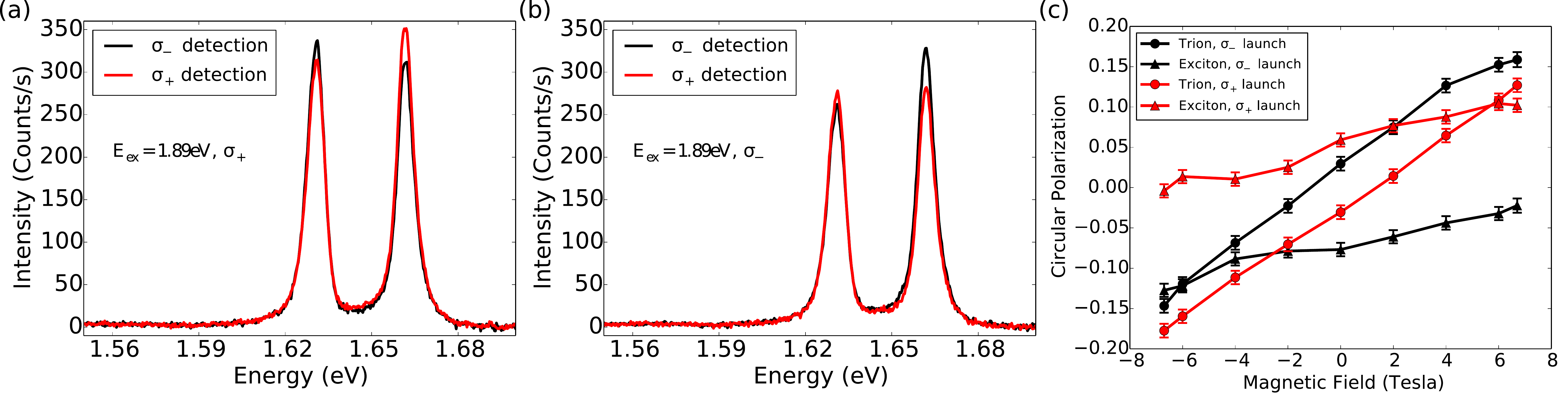}
\end{center}
\caption{(a) Polarization-resolved spectra from D1 taken at zero magnetic field and with $\sigma_+$ excitation, showing $\sigma_+$ polarization of exciton luminescence. (b) Polarization-resolved spectra from D1 taken at zero magnetic field and with $\sigma_-$ excitation. (c) Luminescence polarization versus magnetic field with $\sigma_+$ (red) and $\sigma_-$ (black) excitation for excitons (triangles) and trions (circles). } \label{figs4}
\end{figure*}

Figures \ref{figs4}a and \ref{figs4}b show polarization-resolved luminescence spectra for D1 at $T=4.2$ K and $B = 0$ T taken with $\sigma_+$ and $\sigma_-$ polarized excitation respectively. We observe some preservation of the incident polarization even with our 1.89 eV excitation. We find $P_{\text{exciton}}=\frac{I_{+}-I_{-}}{I_{+}+I_{-}}=6\%$ for $\sigma_+$ excitation  and $P_{\text{exciton}}=-8\%$ for $\sigma_{-}$ excitation indicating $7\%$ average co-polarization of exciton luminescence with the excitation laser. On the other hand, we see counter polarization of $3\%$ for the trion luminescence. We also studied the dependence of the field-induced polarization on excitation handedness: as shown in Fig.\ \ref{figs4}c switching the excitation polarization seemingly adds a constant offset. The small polarization preservation we observe is consistent with studies of polarization preservation in MoS$_2$ using off-resonant excitation \cite{kioseoglou2012valley,mak2012control}. 

\section{2. Background Subtraction and Fitting}

\begin{figure*}[h]
\begin{center}
\includegraphics[width=\textwidth]{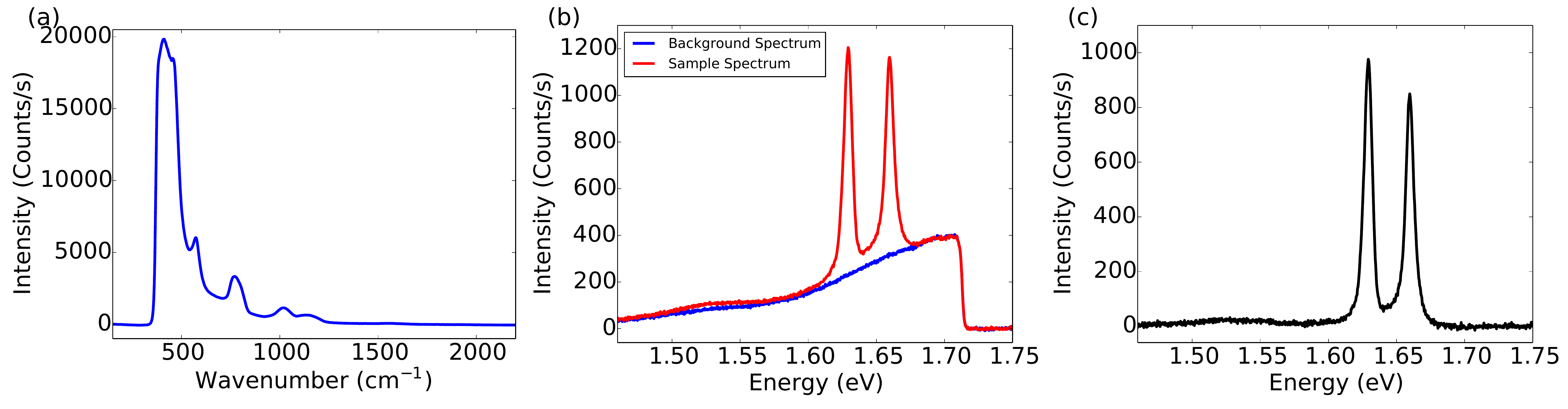}
\end{center}
\caption{(a) Fiber background spectrum excited with 705 nm laser diode, showing fused silica Raman peaks. (b) Comparison of spectra taken with 656 nm excitation laser on the sample (red) and on a nearby region of bare substrate (blue). (c) The result of subtracting the two curves in (b).} \label{figs1}
\end{figure*}

Raman scattering of the excitation laser in the fiber presents a significant background in our experiment, as has been reported elsewhere \cite{hogele2008fiber, ma1996fiber}. A spectrum of fiber Raman excited with 705 nm light is plotted in Fig.\ \ref{figs1}a, showing fused silica Raman peaks \cite{ma1996fiber}. Since we excite with 656 nm light we encounter only the tail of this signal during measurements of MoSe$_2$ luminescence. To account for this background, we take additional spectra with the excitation laser spot on silicon; the background spectrum is then subtracted from the signal after carrying out a dark-count subtraction on both spectra. This is shown in Figs.\ \ref{figs1}b and \ref{figs1}c. In practice, we rescale the background to match the signal spectrum away from the luminescence peaks, to account for laser-power fluctuations and to allow a single background spectrum to be used multiple times. In Figs.\ \ref{figs1}b and \ref{figs1}c we have used the data without rescaling to prove that fiber Raman entirely accounts for the background.

\begin{figure*}[h]
\begin{center}
\includegraphics[width=\textwidth]{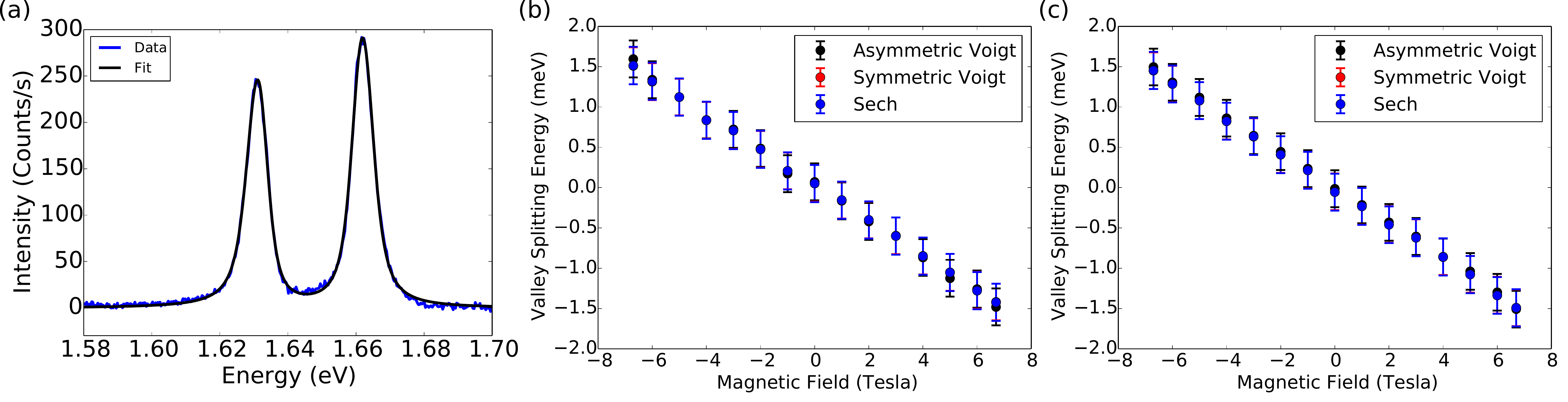}
\end{center}
\caption{(a) Comparison of sample luminescence spectrum (blue) and fit used to locate peak energy (black). The spectrum is fit to the sum of two asymmetric Voigt profiles, with $\chi^2\approx 3$ (b) Trion valley splitting as extracted with fits to asymmetric Voigt (black), symmetric Voigt (red), and hyperbolic secant (blue). (c) Exciton valley splitting as extracted with fits to asymmetric Voigt (black), symmetric Voigt (red), and hyperbolic secant (blue). Valley splittings from asymmetric Voigt fits are presented in Fig.\ 3 of the main text. } \label{figs2} 
\end{figure*}

In the main text we report values for the peak polarization and energy as a function of magnetic field and gating. As described there, we use fits to an asymmetric Voigt profile to extract the peak properties. The Voigt function is defined as:
\begin{equation} 
\frac{1}{\sigma \sqrt{2\pi}}\operatorname{Re}\left\{\exp\left[-\left(\frac{\delta \omega + i\gamma}{\sqrt{2}\sigma}\right)^2\right]\operatorname{erfc}\left[-i\left(\frac{\delta \omega + i\gamma}{\sqrt{2}\sigma}\right) \right]\right\},
\end{equation}
 where $\delta \omega$ is the detuning and $\gamma$ and $\sigma$ are fit parameters characterizing the peak width. As written, the function describes the convolution of a Lorentzian with width $\gamma$ and a Gaussian with width $\sigma$; to make the line shape asymmetric we allow $\gamma$ to take different values for positive and negative detuning. A typical spectrum with fit is plotted in Fig.\ \ref{figs2}a; in this case the $\chi^2$ was about 3. We also tried fitting to other functions, such as a hyperbolic secant and a symmetric Voigt profile. There was no difference in the valley splitting within our errorbars. A comparison of splitting energies between symmetric Voigt, hyperbolic secant and asymmetric Voigt is shown in Figs.\ \ref{figs2}b and  \ref{figs2}c. 

\section{3. Comparison of Data from Multiple Devices}

\begin{figure*}[h]
\begin{center}
\includegraphics[width=\textwidth]{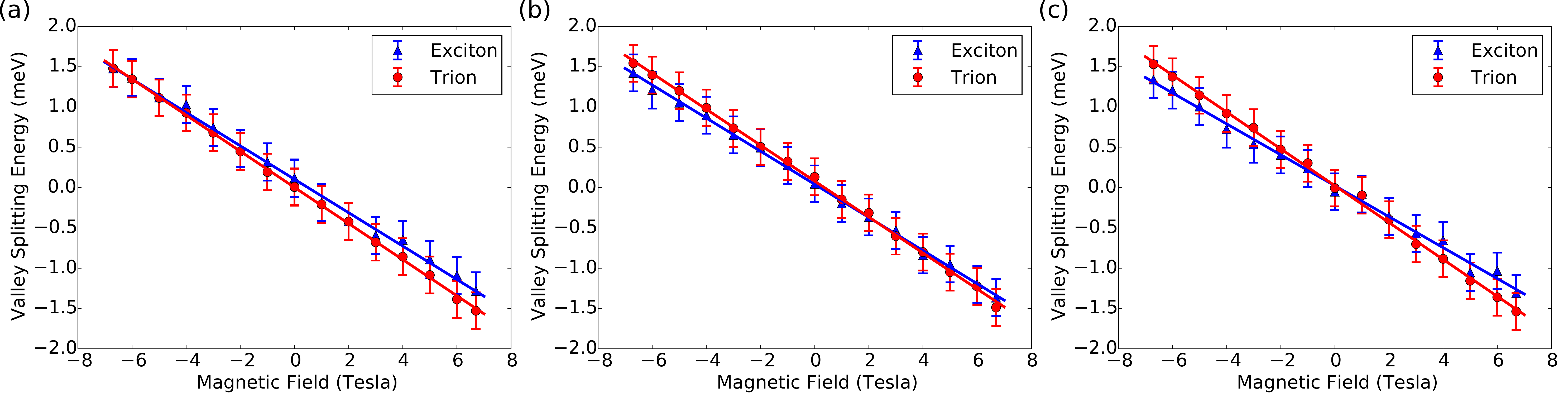}
\end{center}
\caption{(a) Valley splitting data for D1, as defined in the main text. (b) Valley splitting data taken near the center of D3. (c) Valley splitting data taken near one edge of D3.} \label{figs3}
\end{figure*}

We measured the valley splitting versus magnetic field with the back-gate grounded for three different devices. All data were taken at 4.2 K and with 1.89 eV excitation. Valley splitting data not shown in the main text are given in Fig.\ \ref{figs3}; D1 and D2 are defined in the main text, and the additional device is called D3. For D3, we took data at two different positions on the flake. We have also provided Table \ref{tab1} showing the slopes extracted from linear fits to this data. The standard deviation across samples of the trion splitting is 0.003 $\frac{\text{meV}}{\text{T}}$ and the standard deviation of the exciton splitting is 0.01 $\frac{\text{meV}}{\text{T}}$. These values are within the systematic error estimated from the CCD pixel size. For one of the locations on D3, there was a significant discrepancy between the exciton and trion splitting.  

\begin{table}[h]
\begin{center}
    \begin{tabular}{|c| c| c|}
    \hline
     Sample & Exciton Splitting ($\frac{\text{meV}}{\text{T}}$) & Trion Splitting ($\frac{\text{meV}}{\text{T}}$)\\ \hline
   D1 & -0.22 & -0.22 \\ \hline
   D2 & -0.21 & -0.22  \\  \hline
   D3 location 1 & -0.21 & -0.22 \\ \hline
   D3 location 2 & -0.19 & -0.23 \\
   \hline
   
    \end{tabular}
    \caption{Valley splitting for multiple devices in $\frac{\text{meV}}{\text{T}}$, defined as the difference of luminescence peak energies between $\sigma_+$ and $\sigma_-$ polarized light. The error for all values is $\pm 0.01 \frac{\text{meV}}{\text{T}}$. }\label{tab1}
\end{center}

\end{table}

\begin{figure*}[h!]
\begin{center}
\includegraphics[width=\textwidth]{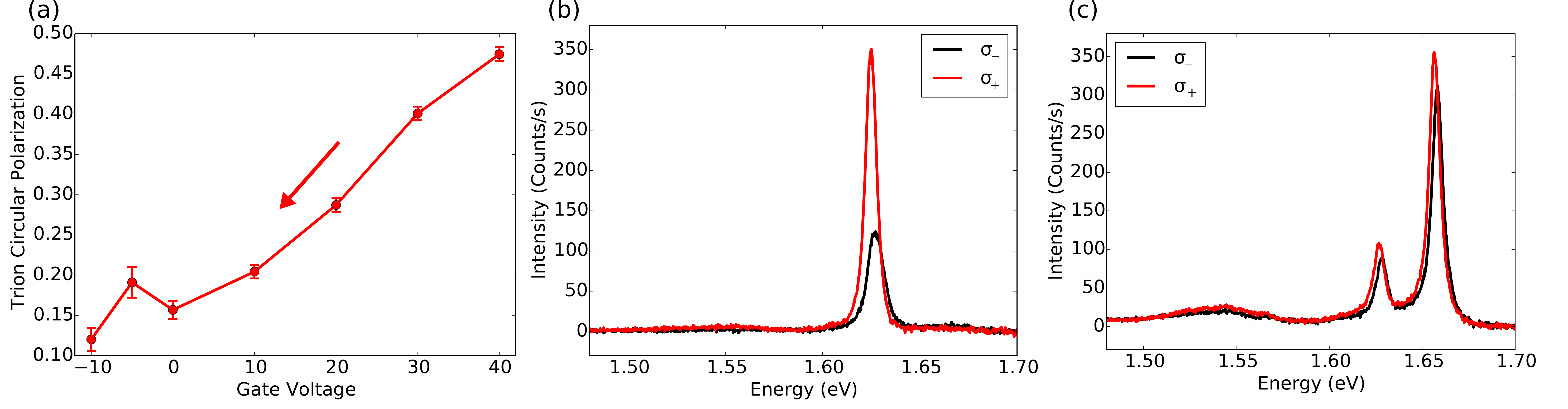}
\end{center}
\caption{(a) Trion peak polarization versus gate voltage at $B=6.7$ T for D4 taken on a downward sweep and using $11 \mu$W excitation with equal intensity in $\sigma_+$ and $\sigma_-$ light (b) Polarization-resolved luminescence spectrum of D4 taken at 6.7 T magnetic field and 40 V back-gate voltage. (c) Polarization-resolved luminescence spectrum of D4 taken at 6.7 T magnetic field and -20 V back-gate voltage. The trion polarization is significantly reduced compared to the 40 V spectrum.} \label{figs7}
\end{figure*}

We also measured the gate dependence of valley splitting and polarization on two devices: D2 and another device not previously defined, D4. The gate dependence of luminescence from D4 is shown in Fig.\ \ref{figs7}. As shown in Fig.\ \ref{figs7}a, for D4 the trion polarization increases from about $10\%$ to over $45\%$ as the electron density is increased. For the data in Fig.\ \ref{figs7} we used excitation light with equal intensity in $\sigma_+$ and $\sigma_-$ polarization, and about 11 $\mu$W excitation power.

\section{4. Further Discussion of the Trion Luminescence and its Gate Voltage Dependence}

\begin{figure*}[h!]
\begin{center}
\includegraphics[width=\textwidth]{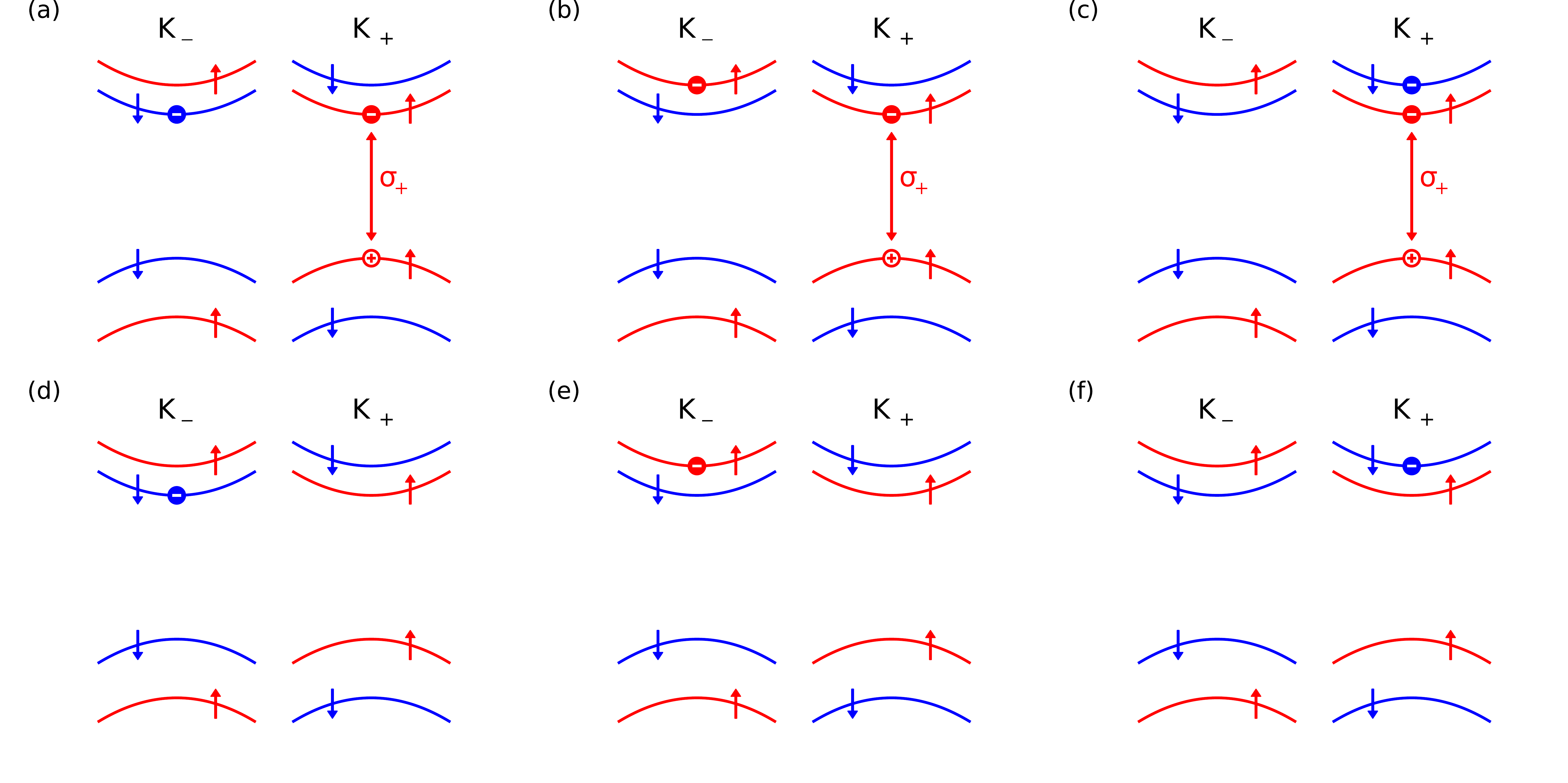}
\end{center}
\caption{Here we show the three possible trion spin-valley configurations which emit $\sigma_+$ polarized light on recombination (panels (a)-(c)) and the corresponding final states after recombination (panels (d)-(f)). In these schematic drawings, the full circles represent the two electrons in the trion, and the open circle represents the hole. We have arranged the panels so that the final state is below the initial state. The configurations shown here are the complete set of trion configurations emitting $\sigma_+$ light, but there are three more which emit $\sigma_-$ light which are related to these via time-reversal symmetry. For the gate-voltage regime considered in our experiment, we expect that  photoluminescence primarily arises from recombination of the trion species in panel (a) and its time-reversed partner.} \label{figs8}
\end{figure*}

Figure \ref{figs8} shows the three possible trion spin-valley configurations which emit $\sigma_+$ polarization light on recombination (upper panels) and the corresponding final states after recombination (lower panels). There are also three more trion configurations not shown in Fig.\ \ref{figs8} which are related to the configurations shown by time-reversal symmetry, and which emit $\sigma_-$ light on recombination. In total there are then six trion configurations expected to have approximately the same binding energy, although the electron-hole exchange interaction is predicted to increase the energy of trions with parallel electron spins by about 6 meV \cite{yu2014dirac}. For MoSe$_2$ at low carrier density, only the lowest conduction bands will be occupied at 4.2 K since the conduction band spin-splitting is predicted to be about 20 meV \cite{PhysRevX.4.011034, PhysRevB.88.085433,PhysRevB.88.245436}. As a result, the trion species in panel (a) is expected to be dominant at low carrier density. Based on Fig.\ 4d of the main text, we see that the conduction band edge for our samples is at approximately $-12$ V on a downsweep and therefore as an upper bound we gate into the conduction band by about $\frac{C\Delta V}{eA}/\frac{m_c}{\pi \hbar^2}\approx 20$ meV at our highest gate voltages (using $C\approx1.2\times10^{-8}$ F/cm$^{-1}$ as the back-gate capacitance per unit area). The presence of trap states means this is probably an overestimate and we expect that the observed luminescence signal primarily arises from recombination of the panel (a) trion (and the time reversed version emitting $\sigma_-$ light) at all gate voltages studied in this work. 

In magnetic field, the total Zeeman energy of the trion can be approximated as the sum of the Zeeman energies of its constituent electrons and hole (the hole Zeeman energy being minus that of the relevant valence band). For example, the photon emitted when the trion in panel (a) recombines has energy: $E_{\text{initial}}-E_{\text{final}}=\epsilon_c+E^{c}_Z-\epsilon_v-E^{v}_Z+\epsilon_c-E^{c}_Z-E_{\text{B}}-\left(\epsilon_c-E^{c}_Z\right)=\epsilon_c-\epsilon_v-E_{\text{B}}+E^{c}_Z-E^{v}_Z=\hbar \omega+E^{c}_Z-E^{v}_Z$, where $E_{\text{B}}$ is the sum of the exciton and trion binding energies (i.e. the total trion binding energy below the electronic band-gap), and $\hbar \omega$ is the trion emission energy for zero magnetic field. The trion valley  splitting is then $2(E^{c}_Z-E^{v}_Z)$ and equal to the exciton valley splitting. Similar calculations give the same results for the transitions shown in panels (b) and (c). 

\begin{figure*}[h!]
\begin{center}
\includegraphics[width=\textwidth]{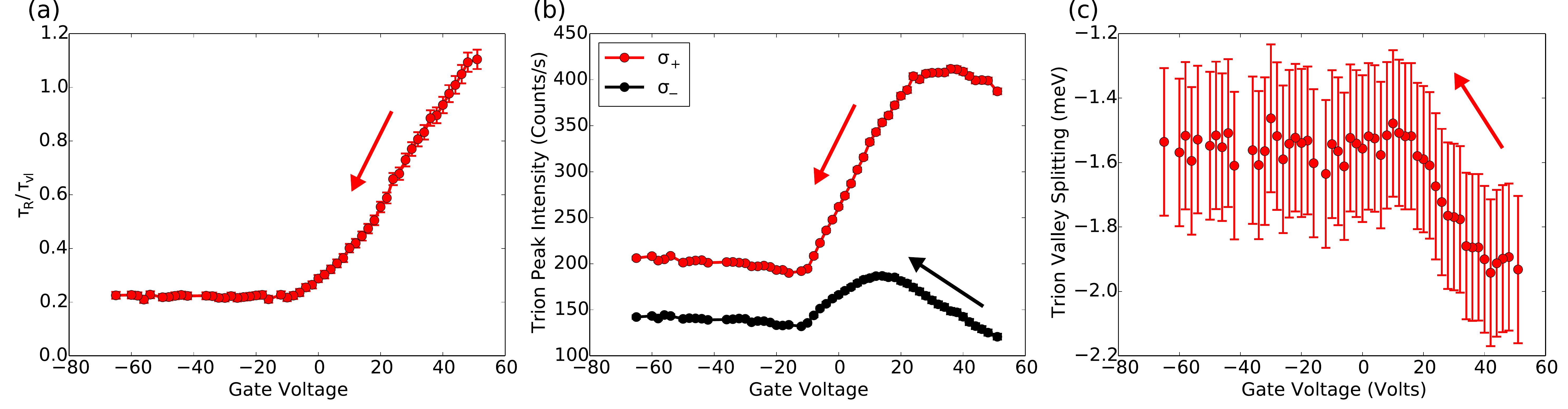}
\end{center}
\caption{(a) Gate voltage dependence of $\frac{\tau_{\text{R}}}{\tau_{\text{vl}}}$ as defined in the text. This is the same dataset as used for Fig.\ 4c of the main text: it was taken for sample D2 at 4.2 K and 6.7 T magnetic field, using an excitation power of about 12.5 $\mu$W and $\sigma_-$ excitation. The arrow represents the direction of the gate-voltage sweep. (b) Peak intensities of trion luminescence in $\sigma_+$ (red) and $\sigma_-$ (black) detection versus gate voltage at 4.2 K and 6.7 T, taken for sample D2. (c)  Trion valley splitting versus gate voltage at 4.2 K and 6.7 T, taken for sample D2.  }  \label{figs9}
\end{figure*}

To estimate the gate dependence of the trion polarization we use a simple rate-equation model. In this model we assume that, for $B>0$, the trion scattering rate from valley $K_+$ to $K_-$ is suppressed by a Boltzmann factor of $e^{-2\beta E^{v}_Z}$ compared to the time-reversed process \cite{PhysRevB.66.235318}, where $\beta=\frac{1}{k_{\text{B}}T}$ with $T$ the effective temperature of the trion population. The argument of the Boltzmann factor is determined by the energy barrier for switching a trion from $K_+$ to $K_-$ valley, which is given by $E_{\text{initial}}(K_-)-E_{\text{initial}}(K_+)=2E^{v}_Z$  for the trion configurations as in panel (a).  We will also assume that, due to our off-resonant excitation, the formation rate $Q$ of $K_+$ and $K_-$ trions is roughly equal. The resulting rate equation is:
\begin{equation}
\begin{cases}
\frac{dn_+}{dt}=Q-n_+/\tau_{\text{R}}+n_-/\tau_{\text{vl}}-n_+e^{-2\beta E^{v}_Z}/\tau_{\text{vl}}\\
\frac{dn_-}{dt}=Q-n_-/\tau_{\text{R}}-n_-/\tau_{\text{vl}}+n_-e^{-2\beta E^{v}_Z }/\tau_{\text{vl}}
\end{cases}
\end{equation}
where $n_{\pm}$ is the trion population in the $K_{\pm}$ valley, $1/\tau_{\text{R}}$ is the trion recombination rate,  and $1/\tau_{\text{vl}}$ is the rate for $K_{-}$ to $K_{+}$ intervalley scattering of the trion. In this simple model we have also ignored the possibility that the recombination rate may depend on the valley. The steady state solution is:
\begin{equation}
P_{\text{trion}}=\frac{n_+-n_-}{n_++n_-}=\frac{\frac{\tau_{\text{R}}}{\tau_{\text{vl}}}\left(1-e^{-2\beta E^{v}_Z }\right)}{1+\frac{\tau_{\text{R}}}{\tau_{\text{vl}}}\left(1+e^{-2\beta E^{v}_Z }\right)}\approx\frac{\frac{\tau_{\text{R}}}{\tau_{\text{vl}}}}{1+\frac{\tau_{\text{R}}}{\tau_{\text{vl}}}}
\end{equation}
where the second equality is obtained by ignoring the Boltzmann factor $e^{-2\beta E^{v}_Z}\approx0.0004$ at 4.2 K. At low gate voltages we find $P_{\text{trion}}\approx0.18$ for the data in Fig.\ 4c of the main text or $\frac{\tau_{\text{R}}}{\tau_{\text{vl}}}\approx0.2$. This is about an order of magnitude larger than the value of $\frac{\tau_{\text{R}}}{\tau_{\text{vl}}}\approx0.03$ found by Ref.\ \cite{PhysRevB.90.075413}; however, their value was obtained in significantly different experimental conditions since they studied WSe$_2$ samples using resonant excitation and at zero magnetic field. In Fig.\ \ref{figs9}a, we plot the intervalley scattering rate normalized to the recombination rate $\frac{\tau_{\text{R}}}{\tau_{\text{vl}}}\approx|P_{\text{trion}}|/\left(1-|P_{\text{trion}}|\right)$ versus gate voltage. The data shows a linear increase in intervalley scattering with carrier density, consistent with the Bir-Aronov-Pikus mechanism for intervalley hole scattering by the background conduction electrons \cite{mak2012control, bir1975spin}. As discussed in section 7 of the supplement, we also observe a decrease in the trion valley polarization with increasing excitation power. This is qualitatively consistent with the rate-equation model assuming that the effective temperature of the trion population increases with excitation power. 

In Figure \ref{figs9}b we show the peak intensity of trion luminescence in $\sigma_+$ and $\sigma_-$ detection versus gate voltage. At small gate voltages, the trion luminescence intensity increases with increasing gate voltage for both $\sigma_+$ and $\sigma_-$ detection, but the intensity of $\sigma_-$ luminescence begins to decrease significantly above 15 V.  As shown in Fig.\ \ref{figs9}c, the trion valley splitting changes only for gate voltages above about 15 V, suggesting that the increase in the valley splitting magnitude and decrease in the $\sigma_-$ intensity could be related. Since the trion can recombine with a range of center of mass wavevectors \cite{ross2013electrical}, the change in valley splitting may result from a change in the $k$-space distribution of the $K_-$ trion population rather than a change in the band Zeeman energies; the trion recombination energy redshifts as the final state center of mass momentum is increased. The redshift of $K_+$ trions compared to $K_-$ trions would then correspond to a hot luminescence effect where  the trion scattering rate from the $K_-$ to $K_+$ valley is larger at larger wavevectors.

\section{5. Calculations of the Exciton Valley Splitting within $\mathbf{k\cdot p}$ Theory and Tight-binding Models}

The magnetic moment of a band $b$ can be calculated in $\mathbf{k\cdot p}$ theory using the formula \cite{ yafet1963solid, PhysRevB.77.235406, PhysRev.102.1030, PhysRev.118.1534, wu2013electrical,PhysRevX.4.011034}:
\begin{equation}\label{eqn1}
m_b=-\frac{\mu_{\text{B}}}{2  m_e}\sum_{a\neq b}\frac{|P^{ba}_+|^2-|P^{ba}_-|^2}{\epsilon_b-\epsilon_a}
\end{equation}
where $m_e$ is the free electron mass, $\mu_{\text{B}}$ is the Bohr magneton, $\epsilon_a$ is the energy of band $a$, and $P^{ba}_{\pm}=\left \langle b \right|p_x\pm i p_y \left|a\right \rangle$ is proportional to the optical matrix element for $\sigma_{\pm}$ light between Bloch states $\left|a\right \rangle$ and $\left|b\right \rangle$. The formula above gives the $z$ component of the magnetic moment, assuming $\{ x,y,z\}$ form a right-handed coordinate system. As discussed in early papers on Bloch electrons in magnetic fields \cite{yafet1963solid, PhysRev.102.1030}, this formula includes both contributions from the phase winding of the Bloch state within a unit cell (the atomic contribution to the magnetic moment) and the phase winding on the scale of multiple unit cells (the intercellular contribution to the magnetic moment). The optical matrix elements are determined by the $\mathbf{k\cdot p}$ Hamiltonian matrix elements, since $H_{\mathbf{k\cdot p}}=\frac{\hbar}{2 m_e}\left(k_+p_-+k_-p_+\right)$, where $k_{\pm}=k_x\pm ik_y$, $p_{\pm}=p_x\pm ip_y$, and where $\mathbf{k}$ and $\mathbf{p}$ are the wavevector and momentum operator respectively. 

In Table V of Ref.\ \cite{PhysRevX.4.011034}, Korm\'anyos \textit{et al.} give the non-zero $\mathbf{k \cdot p}$ matrix elements within their theory. The resulting valley Zeeman energies (as defined in the main text) are:
\begin{equation}
E_Z^c/\mu_{\text{B}}=\frac{2 m_eB |\gamma_3/\hbar|^2}{\epsilon_c-\epsilon_v}-\frac{2 m_e B|\gamma_5/\hbar|^2}{\epsilon_c-\epsilon_{v-3}}-\frac{2 m_e B|\gamma_6/\hbar|^2}{\epsilon_c-\epsilon_{c+2}}
\end{equation}
in the conduction band and
\begin{equation}
E_Z^v/\mu_{\text{B}}=-\frac{2 m_e B |\gamma_3/\hbar|^2}{\epsilon_v-\epsilon_c}+\frac{2 m_e B|\gamma_2/\hbar|^2}{\epsilon_v-\epsilon_{v-3}}+\frac{2 m_e B|\gamma_4/\hbar|^2}{\epsilon_v-\epsilon_{c+2}}
\end{equation}
in the valence band, where $\epsilon_{c+2}$ is the energy  of the second band above the conduction band and $\epsilon_{v-3}$ is the energy of the third band below the valence band. Here the parameters $\gamma_i$ are related to the interband optical matrix elements, and the authors of Ref.\ \cite{PhysRevX.4.011034} determine relevant combinations of these parameters via fits to the DFT band structure; details of the fitting procedure can be found in Refs.\ \cite{PhysRevB.88.045416, ModelHamiltoniansandmaterialparameters}. In our case, the precise values of these parameters are not important, as we will focus on the relationship between the effective masses and valley splitting that can be derived using the $\mathbf{k \cdot p}$ approach. The $\mathbf{k \cdot p}$ theory effective masses can be written in terms of the $\gamma_i$ similar to the Zeeman splitting (see Eq. B6 of Ref.\ \cite{PhysRevX.4.011034}). Some simple algebra then allows us to obtain:
\begin{equation}
g^{\text{vl}}_{\text{ex}}=\frac{2(E_Z^c-E_Z^v)}{\mu_{\text{B}}}=4-2\left(\frac{m_e}{m_c}-\frac{m_e}{|m_v|}\right)
\end{equation}
where $m_{c(v)}$ is the effective mass of the conduction (valence) band. As long as the effective masses for conduction and valence band are approximately equal, as expected from first principles calculations \cite{PhysRevB.86.115409, ModelHamiltoniansandmaterialparameters}, the valley splitting calculated this way will be close to $g^{\text{vl}}_{\text{ex}}=4$ and have the opposite sign to our measurements. For example, taking $m_c=0.49m_e$ and $|m_v|=0.59m_e$ (these values are from \cite{ModelHamiltoniansandmaterialparameters}) gives $g^{\text{vl}}_{\text{ex}}=3.3$. 

The exciton valley splitting can also be calculated using a lattice model. For example, Ref. \cite{PhysRevLett.108.196802} originally proposed a model Hamiltonian for TMDs based on hybridization of $d$-orbitals at different Mo lattice sites. Such a lattice model neglects the atomic-scale structure of the wave function, and therefore the Zeeman coupling to the $d$-orbital magnetic moment must be introduced by hand \cite{srivastava2014valley, aivazian2014magnetic}. This gives a contribution to the band Zeeman energies of $E^{c}_{Z,\text{a}}=0$ and $E^{v}_{Z,\text{a}}=2\mu_{\text{B}}$, as discussed in the main text. Aside from this contribution, there is the magnetic moment due to phase winding of the Bloch states on the intercellular scale. This quantity can be calculated using the $\mathbf{k \cdot p}$ theory formula Eq.\ \ref{eqn1} above, but this time within the reduced Hilbert space of the lattice model. For example, the Hamiltonian in the massive Dirac fermion model is:
\begin{equation}
H=
\begin{pmatrix}
\epsilon_c & \tau \gamma_3 q_{{-\tau}}  \\
 \tau\gamma^{\ast}_3 q_{\tau} & \epsilon_v \\
\end{pmatrix}
\end{equation} 
written in the basis of band-edge Bloch functions $\left\{ \left|c\right\rangle, \left|v\right\rangle\right\}$. The resulting value for the intercellular Zeeman energy is $E_{Z,\text{ic}}^{c(v)}=\mu_{\text{B}}\frac{2 m_eB |\gamma_3/\hbar|^2}{\epsilon_c-\epsilon_v}$. Here we have used that $\left\langle c|p_+|v\right\rangle=2m_e \gamma_3/\hbar$. We note that $\frac{\hbar^2}{2m_c}=\frac{|\gamma_3|^2}{\epsilon_c-\epsilon_v}$ for this model so that the Zeeman energy is simply $\mu_{\text{B}}B \frac{m_e}{m_c}$. In a given valley this contribution shifts the energy levels in the conduction and valence bands in the same way, and therefore does not contribute to the exciton valley splitting. The total exciton valley splitting for this model is  $2(E^c_{Z}-E^v_Z)=2(E^c_{Z,\text{a}}-E^v_{Z,\text{a}})+2(E^c_{Z,\text{ic}}-E^v_{Z,\text{ic}})=-4\mu_{\text{B}}$ as discussed in the main text. The same approach of separately treating the inter and intra cellular contributions can be used to calculate the exciton valley splitting in more general lattice models where the electron and hole masses are not equal, giving a value for the exciton valley splitting which differs from the bare $d$-orbital one \cite{srivastava2014valley, aivazian2014magnetic}. 

Finally, we discuss the effective Hamiltonian for excitons in magnetic field. The exciton Hamiltonian is found by subtracting the conduction and valence band dispersions and adding the electron-hole Coulomb interaction $V$:
\begin{align}
H_{\text{ex}} & =H_{c}\left(-i\hbar\nabla_e,\mathbf{r}_e\right)-H_{v}\left(i\hbar \nabla_h,\mathbf{r}_h \right)+V\left(|\mathbf{r}_e-\mathbf{r}_h|\right)\\
& =\frac{\hbar^2}{2m_c} \left(-i\hbar\nabla_e+e\mathbf{A}(\mathbf{r_e})\right)^2-\frac{\hbar^2}{2m_v} \left(-i\hbar\nabla_h-e\mathbf{A}(\mathbf{r_h})\right)^2+V\left(|\mathbf{r}_e-\mathbf{r}_h|\right)+\frac{1}{2}g^{\text{vl}}_{\text{ex}}\mu_{\text{B}}B \tau.
\end{align}

Following Refs. \cite{knox1963theory, gippius1998excitons}, we carry out a gauge transformation to find a one-body Hamiltonian for excitons with zero center of mass momentum:
\begin{equation}
H_{\text{ex}}^{\tau}=\frac{\hbar^2}{2\mu} \mathbf{k}^2+\frac{\hbar e B}{2}\left(\frac{1}{m_c}-\frac{1}{|m_v|}\right)l_z+\frac{e^2B^2}{8\mu}\mathbf{r}^2+V\left(|\mathbf{r}|\right)+\frac{1}{2}g^{\text{vl}}_{\text{ex}}\mu_{\text{B}}B \tau \label{eq:exciton-Ham}
\end{equation}
where $\mathbf{r}=\mathbf{r}_e-\mathbf{r}_h$ is the electron-hole separation, $\mathbf{p}$ is the associated canonical momentum,  $\mu=m_c|m_v|/(m_c+|m_v|)$, and $l_z=\hat{z}\cdot\left(\mathbf{r}\times \mathbf{p}\right)$. For bright  excitons we assume $l_z=0$, i.e. that they are  $s$-type \cite{ PhysRevB.89.125309, PhysRevB.89.205436,PhysRevB.88.045318}. Therefore the only term which can give rise to a linear magnetic field dependence of the exciton energy is the last term in Eq.\ \ref{eq:exciton-Ham}, which describes a Zeeman-like coupling of the exciton valley degree of freedom to the  magnetic field. 

We also estimate the energy shift due to the quadratic term $\frac{e^2B^2}{8\mu}\mathbf{r}^2$ in the exciton Hamiltonian. In the regime where the magnetic length ($l_B=\sqrt{\frac{\hbar}{e B}}$) is smaller than the exciton Bohr radius, this term leads to a quadratic shift of the exciton transition energy as demonstrated in experiments on quantum wells  \cite{gippius1998excitons,PhysRevB.48.8848,PhysRevB.43.4152,PhysRevB.34.4002}. Theoretically, this could manifest in our experiments as a quadratic term in the valley-averaged transition energy, but due to the small exciton Bohr radius for TMDs (1-3 nm \cite{PhysRevB.88.045318,PhysRevB.89.205436}) the correction should be small. We can estimate the diamagnetic shift using perturbation theory with the Wannier model above: the result is a quadratic increase of order $\frac{1}{8}\hbar(\omega_c+\omega_v)\left(\frac{a_{\text{B}}}{l_B}\right)^2\approx 7\, \mu \text{eV}$ at 6.7 T, where $\omega_{c(v)}$ is the electron (hole) cyclotron frequency, and $a_{\text{B}}$ is the exciton Bohr radius. This energy shift is below our measurement sensitivity.

\section{6. Experimental Determination of the Sign of the Valley Splitting}

In the main text, we define the valley splitting as the difference of peak luminescence energies between $\sigma_+$ and $\sigma_-$ polarized emission. Furthermore, $\sigma_{\pm}$ polarization is defined as the circular polarization which carries $\pm\hbar$ angular momentum per photon along the field direction for $B>0$. Equivalently, $\sigma_+$ ($\sigma_-$) polarized light can be defined as the light with electric field vector rotating counter-clockwise (clockwise) in time around the positive $B$ axis. The convention for $B>0$ is defined in Fig.\ 1a of the main text. To determine the sign of the splitting, we used two methods. 

First, we determined the rotational settings of the detection polarizer corresponding to different circular polarizations of emission. To do this, we launched circularly-polarized laser light into the cryostat objective lens from the sample space, and found the settings of the detection polarizer which maximized the resulting signal. The circularly-polarized light was generated by sending linearly polarized light through a $\lambda/4$ plate with the light polarized at $45^{\circ}$ to the waveplate axes. Given knowledge of the waveplate axes and their orientation relative to the light polarization, the handedness of circularly-polarized light produced in this fashion can be determined. We also checked the assignment of the waveplate fast and slow axes by shining circularly-polarized light of a known handedness through the waveplate and analyzing the resulting linear polarization. For this test, the circularly-polarized light was generated using two N-BK7 prisms in a Fresnel rhomb geometry, so that the resulting handedness could be determined from the Fresnel equations. We determined the field direction using a calibrated Hall probe. The considerations above determine the rotational settings of the detection polarizer corresponding to detection of $\sigma_+$ and $\sigma_-$ emission.

We also compared the valley splitting for MoSe$_2$ to magnetoluminescence measurements for a (110) cut, undoped, $p$-type CdTe substrate (from MTI Corporation). For $p$-type CdTe, the acceptor-bound exciton luminescence shows a four-fold splitting under magnetic field applied in the Faraday geometry. The optical selection rules lead to circular polarization of these peaks, so that two are $\sigma_+$ polarized and two are $\sigma_-$ polarized.  With the detection polarization determined as discussed above, we find peak splitting and selection rules for CdTe in agreement with those found by Refs. \cite{malyavkin1981bound,molva1983magneto,molva1985magneto}. In particular, given that the lowest energy acceptor-bound exciton luminescence peak for CdTe is $\sigma_-$ polarized (for $B>0$), we know that the lowest energy MoSe$_2$ peak indeed originates from $\sigma_+$ polarized luminescence (for $B>0$) as indicated in the main text.

\section{7. Power Dependence of Trion polarization}

\begin{figure*}[h]
\begin{center}
\includegraphics[width=\textwidth]{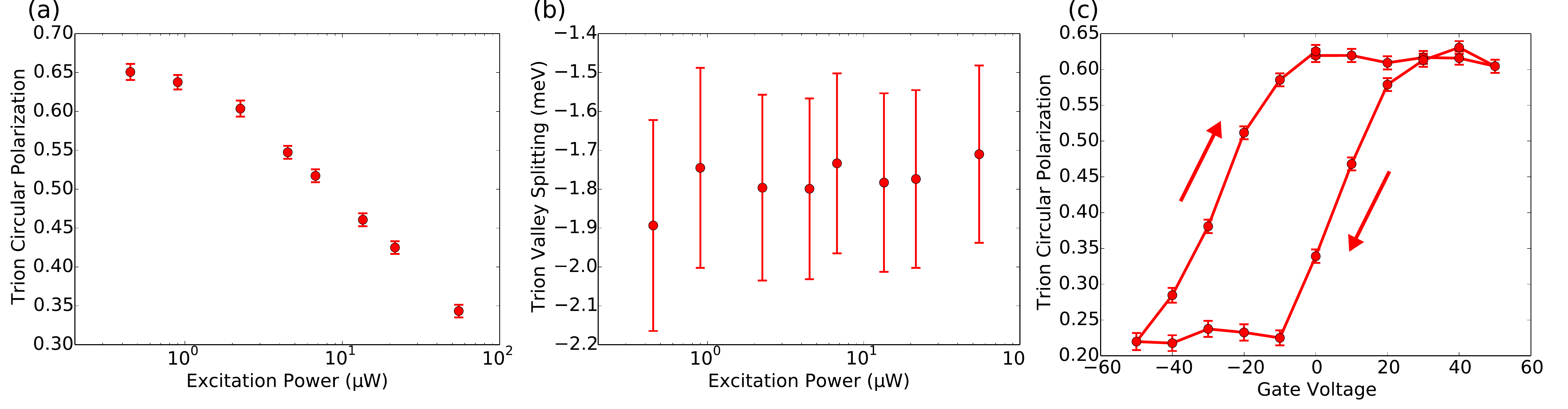}
\end{center}
\caption{(a) Trion peak circular polarization versus power in the $n$-type regime, for $B = 6.7$ T, and excited with $\sigma_-$ polarized light.  (b)  Trion valley splitting versus power in the $n$-type regime and for $B = 6.7$ T. (c) Trion peak circular polarization versus gate voltage, taken at 6.7 T and using about 1.1 $\mu$W excitation power. } \label{figs6}
\end{figure*}

As shown in Fig.\ \ref{figs6}a, the trion luminescence polarization increases to about $65\%$ circularly-polarized as the power is reduced for $B=6.7$ T, $T=4.2$ K, and in the regime of high electron density. On the other hand, we see no power dependence of the trion peak splitting (see Fig.\ \ref{figs6}b). Within our rate equation model, the power dependence of trion polarization arises from changes in the lattice temperature, or the effective temperature of the trion population which may not be equilibrium with the lattice. A thermometer mounted on the chip holder shows $<50$ mK sample heating under more than 200 $\mu $W excitation, suggesting that the lattice heating is small. Figure \ref{figs6}c shows the gate dependence of trion polarization at 6.7 T and 4.2 K, with an excitation power of about 1.1 $\mu$W; the fractional increase in the trion polarization with gate voltage is similar to data shown in the main text (taken with about 11 $\mu$W excitation).

\pagebreak

\end{document}